\documentclass[12pt]{elsarticle}

\usepackage{amssymb}
\usepackage{float}
\usepackage{xcolor}
\usepackage{amsmath}

\begin{document}

\begin{frontmatter}



\author{Mamata Das, Selvakumar K., P.J.A. Alphonse\corref{cor1}}
\ead{dasmamata.india@gmail.com, kselvakumar@nitt.edu, alphonse@nitt.edu}

\affiliation{organization={NIT Trichy},
             city={Tiruchirappalli},
             postcode={620015},
             state={Tamil Nadu},
             country={India}}

\title{Identifying Essential Hub Genes and Protein Complexes in Malaria GO Data using Semantic Similarity Measures}






\begin{abstract}
Hub genes play an essential role in biological systems because of their interaction with other genes. A vocabulary used in bioinformatics called Gene Ontology (GO) describes how genes and proteins operate. This flexible ontology illustrates the operation of molecular, biological, and cellular processes ($\mathbb{P}_{mol}, \mathbb{P}_{bio}, \mathbb{P}_{cel}$). There are various methodologies that can be analyzed to determine semantic similarity. Research in this study, we employ the jackknife method by taking into account $4$ well-liked Semantic similarity measures namely Jaccard similarity, Cosine similarity, Pairsewise document similarity, and Levenshtein distance. Based on these similarity values, the protein-protein interaction network (PPI) of Malaria GO (Gene Ontology) data is built, which causes clusters of identical or related protein complexes ($P_x$) to form. The hub nodes of the network are these necessary proteins. We use a variety of centrality measures to establish clusters of these networks in order to determine which node is the most important. The clusters' unique formation makes it simple to determine which class of $P_x$ they are allied to.
\end{abstract}




\end{frontmatter}



\section{Introduction}
\label{sec:sample1}
Genes are segments of deoxyribonucleic acid (DNA) that contain instructions for encoding particular proteins. The actions that proteins take are governed by genes. Gene annotation is the process of identifying the coding areas and functionality of genes. It aids in locating structural components and connecting their corresponding function to each gene's location. It contains every required biological information to create any kind of living thing. Gene annotation enables the identification and prediction of $P_x$ functions, enabling additional comparative research \cite{he2016evolutionary}. A crucial role in biological processes is played by proteins. Understanding $P_x$ is beneficial for understanding how an organism is put together, but it also aids in disease prediction and the identification of potential target cells \cite{spirin2003protein}\cite{baruri2022empirical}. Essential proteins are crucial for preserving cellular life. 
The largest PPI network's (PPIN) topology has been found to have a property \cite{barabasi1999emergence}, which means that the nodal degree distributions of the network are power-law distributions (very close to the cutoff points) \cite{barabasi2004network, yook2004functional}. The result is that degrees are not scaled according to any specific scale. Nonetheless, researchers studying PPINs have long established an arbitrary cutoff point above which all proteins with degrees greater than this cutoff are considered to be uniquely unique and are called hub proteins ($P_{hub}$) \cite{jeong2001lethality, he2006hubs}. While $P_{hub}$ are arbitrarily defined, they often have unique biological characteristics, making them appealing for the introduction. $P_{hub}$ play a key role in forming a modular protein interaction network and, as some studies suggest, may also be more evolutionarily conserved than non-hub proteins ($P_{\neg hub}$) \cite{albert2000error,han2004evidence}. They are therefore frequently found to be more crucial than $P_{\neg hub}$.
It is possible to refer to it as a biological lexicon that illustrates the functions of the $3$ major divisions of $\mathbb{P}_{mol}$, $\mathbb{P}_{bio}$, and  $\mathbb{P}_{cel}$. If there is a resemblance between two statements, it is assumed that both of the sentences express the same meaning. Similarity between two sentences is defined based on the structure and syntax of the sentences. We perform semantic similarity checks on GO phrases in order to assign a numerical value (a ``measuring value") to the GO term. Similarity approaches can be used to locate related genes based on their functions using the ontological data resource (nominal data) \cite{ayllon2018new}.

Jaccard's similarity, Cosine similarity,  pairwise similarity, Levenshtein, based on measurements of distance and ratio, and are some of the popular text similarity metrics. Most of these measures look for terms that are used frequently in both texts in order to quantify how important they are to the sentence.
These measures often yield results in the form of a numeric value between $0$ and $1$. These metrics enable us to compare two texts and determine how similar the phrases are. The PPIN, which may be either directed or undirected, is a graphical representation of proteins related to one another by edges. Typically, these networks show a biological process. 
To identify the node in a network that has the most influence, centrality measurements are used. Degree centrality(DC), closeness centrality(CC), betweenness centrality(BC), eigenvector centrality(EC), harmonic centrality, second order centrality, group centrality, and many more are examples of centrality measurements. These metrics provide each node in a network with a numerical number to indicate their contribution to the network. The central node is the node with the highest centrality value and is therefore connected to the majority of the network's nodes. The biological purpose of the hub node is specified by the PPIN, where hub nodes are treated as clusters. By placing the data in a certain group or category to which it belongs, clustering aids in data analysis. 
The clustering coefficient can be used to determine the degree to which nodes in a network cluster together. The average clustering across the entire network is shown via average coefficient clustering, along with the cluster's degree of completion. 
The creation of clusters of identical or related $P_x$ is suggested in this research. We applied and assessed multiple similarity algorithms to discover the best one to detect related genes ($P_x$) using the gene annotation dataset as the training dataset, which comprises gene names and their capabilities or behaviors. We create the PPI network based on the outcomes of related genes. We utilised and examined multiple centrality measures to build clusters of these networks in order to determine the hub node, or the node with the greatest influence. The clusters' unique formation makes it simple to determine which class of $P_x$ they belong to. Let $V$ be the set of vertices and $E$ be the set of edges in the graph $\mathit{G(V,E)}$. When discussing a general graph, we use the terms vertices or nodes and edges, and when discussing a protein interaction network, we use the terms protein and interaction.
Proteins, nucleic acids, and tiny molecules are necessary for a cell to build a dense network of molecular interactions. Nodes and edges are how molecules interact with one another. The molecular structure's network architecture provides information on the organisation and function of proteins. Protein clusters are created when strongly linked nodes interact to build protein networks \cite{spirin2003protein}. In 2007, \cite{wang2007new} suggested a fresh approach in which an algorithm is put into practise to ascertain how semantically comparable GO concepts are. using this approach to assess the functional similarity of genes. Utilizing online-based methods for gene similarity, results of gene grouping based on similarity values are obtained. We also looked at \cite{ayllon2018new}, where they assessed how advances in genomics research and technology have revealed the dynamic structure and function of genes. Genome annotation is the process of identifying genetic components and their purpose. It is possible to store this in text format. We can thus investigate the query or view the genomics data as a result. 

Comparable gene expression patterns suggest that the biological functions of the genes are likely to be similar. Bringing together genes with similar functions is the basic goal of clustering. 
An ontological annotation of data resources serves as the foundation for a semantic similarity metric \cite{lord2003investigating}. Data and annotations can be found in several bioinformatics resources. 
The foundation study \cite{he2016evolutionary} offers a method for identifying $P_x$. Given how complex the structure is, there is an algorithm for estimating $P_x$, and considerable work has gone into its creation. With the help of a probabilistic Bayesian Network (BN), they create the complex subgraph. The parameters of the BN model are learned using training sets of well-known complexes. It extracts the traits that are used to separate complicated objects from simpler ones. This experiment demonstrates that EGCPI can detect $P_x$ more accurately when utilising evolutionary graph clustering. A research \cite{kang2011centralities} that was published in $2011$ made the suggestion that node centrality measurements, which are crucial for a variety of graph applications and biological network analysis, should be used instead. Many different approaches for defining centrality have been proposed, from the most basic (like node degree) to the most complex and scalable. For evaluating the relative significance of nodes within a graph of small nodes, centrality is frequently used. The many centrality measurements include eigenvector centrality, betweenness centrality, near centrality, and centrality degree. A article \cite{zhang2016detecting} for predicting crucial proteins by combining network topology for cell function was published in 2018 as well. Network levels matter based on PPIs. GO similarity measurements and centrality approaches are used to identify important proteins on PPI networks.

The relationship between hub proteins and essentiality in the S. cerevisiae physical interaction network has been explored, but the characteristics of essential modules and the differences in topological properties between essential and non-essential proteins are not well understood. In \cite{song2013hub}, they have found that essentiality is a modular property, with the number of intra-complex or intra-process interactions being a better predictor of essentiality than overall interaction count. Furthermore, essential proteins within essential complexes have a higher number of interactions, particularly within the complex itself. Identifying key proteins from PPI networks is crucial, but high false positive rates hinder current computational methods. In paper \cite{xue2023comparative}, they have proposed a strategy to construct reliable PPI networks by using Gene Ontology (GO)-based semantic similarity measurements. Author have calculated confidence scores for protein pairs under three annotation terms namely MF, BP, and CC (Molecular function, Biological process, and Cellular component) using five semantic similarity metrics (Jiang, Lin, Rel, Resnik, and Wang). Low-confidence links are filtered out, resulting in refined PPI networks. Six centrality methods are applied and compared, showing that the performance under refined networks is better than under the original networks. Among the metrics, Resnik with a BP annotation term performs the best, highlighting its favorable choice for measuring the reliability of protein links. PPIs are crucial for cellular processes, and hubs play a vital role in maintaining the structure of protein interaction networks. \cite{kenley2011differentiating} This study introduces a novel measure to identify and differentiate two types of hubs, party hubs and date hubs, based on semantic similarity and Gene Ontology data. By combining this measure with centrality measures, the study demonstrates accurate detection of potential party hubs and date hubs, matching confirmed hubs with high accuracy. In the field of molecular biology, identifyin PPIs is crucial. While experimental methods have limitations, computational approaches using semantic similarity from GO annotation have gained attention. \cite{zhang2016protein} study proposed a GO-based method for predicting protein-protein interactions by integrating different similarity measures derived from the GO graph structure. By combining information from both the ascending and descending parts of the three ontologies, the method achieved the best performance in predicting PPIs, demonstrating its effectiveness in this area.
\section{Materials and methods}\label{sec3}
\begin{figure}[H]
	\begin{minipage}[b]{0.45\linewidth}
		\centering
		\includegraphics[width=4cm, height=4cm]{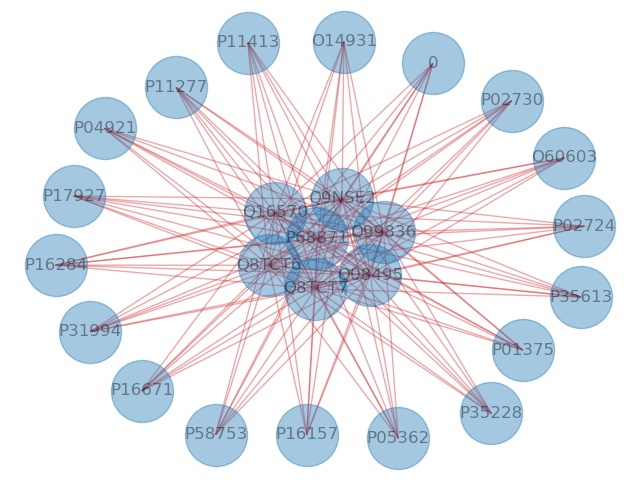}
		\caption{PPIN from ${similarity}_j$}
		\label{random_jaccard_ppi}
	\end{minipage}
	\begin{minipage}[b]{0.45\linewidth}
		\centering
		\includegraphics[width=4cm, height=4cm]{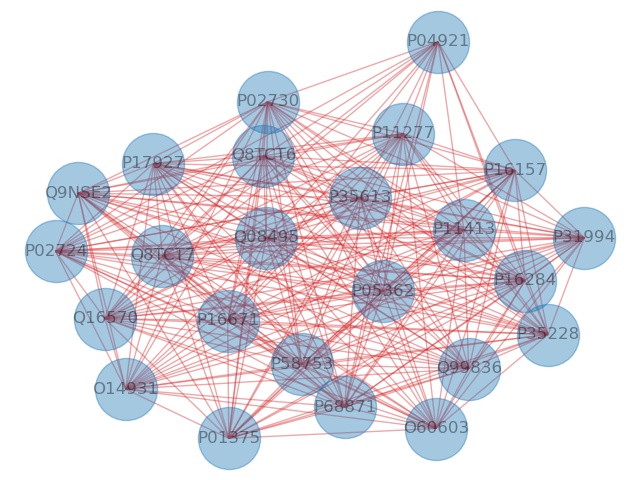}
		\caption{PPIN from ${similarity}_c$}
		\label{random_cs_ppi}
	\end{minipage}
	\begin{minipage}[b]{0.45\linewidth}
		\centering
		\includegraphics[width=4cm, height=3.8cm]{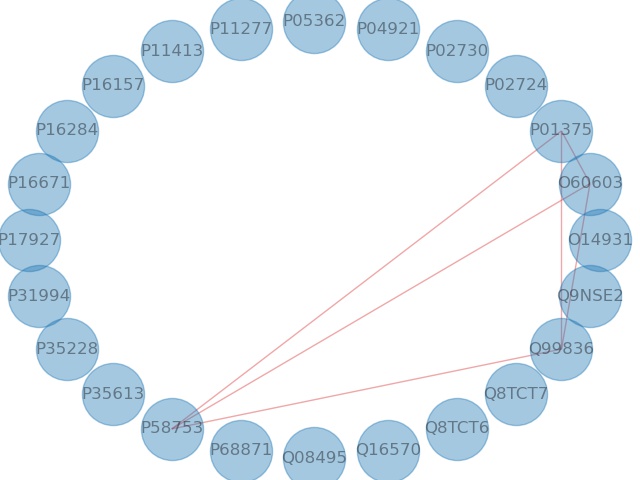}
		\caption{PPIN from ${similarity}_p$ }
		\label{random_ps_ppi}
	\end{minipage}
	\hspace{5mm}
	\begin{minipage}[b]{0.45\linewidth}
		\centering
		\includegraphics[width=4cm, height=4cm]{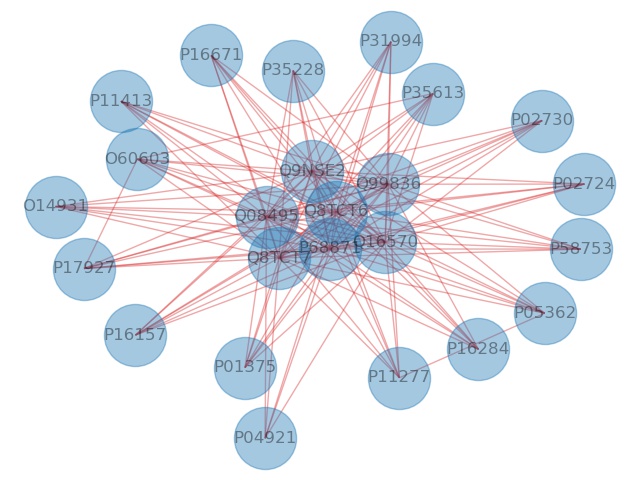}
		\caption{PPIN from ${similarity}_l$}
		\label{random_l_ppi}
	\end{minipage}
\end{figure}

We have used $24$ Malaria GO data \cite{das2021markov}. The data has been taken from UniProtKB/Swiss-Prot database which are reviewed \cite{uniprot2023uniprot}. The used dataset has been mentioned in Table \ref{tab1} where term Entry define Unique and stable entry identifier and Gene Names define Name(s) of the gene(s) encoding the protein. We have done similarity analysis on the data and get the similrity matrix. The edge information (we can say protein interaction data) get from similarity matrix. We have created the PPI network from these edge information. In the PPIN, the nodes represent proteins and edges denote biological interactions between protein pairs. The global properties of each PPIN has been mentioned in Table \ref{tab3}. We may see the similarity matrix in Fig. \ref{hub_js} to \ref{hub_l} where cosine similarity contained maximum no. of $1$.  Fig. \ref{random_cs_ppi} to \ref{random_l_ppi} are showing the PPI network genarated from four similarity measure. We have used Networkx, a well-liked software programme written in Python, to create a PPI network. Here, the PPI networks are undirected. As cosine similarity matrix is containing maximum $1$ values, we will measure the centrality score for this network only. We have chossen four most important centrality score measure DC, CC, BC, EC with additional PR features. The centrality scores are showed in \ref{tab4}.The threshold values for each category were derived by averaging the values of each centrality measure ($th_{value}$). By using the $th_[value]$ we can locate the node that acts as the hub in the network. The centrality score $< th_{value}$ of any protein is less influential than the protein's own centrality score $\ge th_{value}$. The hub protein was then identified by getting the intersection of all the significant proteins in each category (DC, CC, BC, EC, and PR).
\begin{table}[htbp]%
	\begin{minipage}{\textwidth}
		\caption{\small $24$ Malaria Gene name}\label{tab1}%
		\begin{tabular}{@{}llll@{}}
  \hline
			Entry & Gene Names & Entry & Gene Names  \\
   \hline
			
			\small P58753  & \small  {TIRAP MAL}  & \small  P16671  & \small  CD36 GP3B GP4\\
			\small O60603  & \small  TLR2 TIL4  & \small  P04921  & \small  GYPC GLPC GPC\\
			\small Q9NSE2  & \small  CISH G18  & \small  P31994  & \small  FCGR2B CD32 FCG2 IGFR2\\
			\small P02724  & \small  GYPA GPA  & \small  P35228  & \small  NOS2 NOS2A\\
			\small P68871  & \small  HBB  & \small  P35613  & \small  BSG UNQ6505/PRO21383\\
			\small P17927  & \small  CR1 C3BR  & \small  O14931  & \small  NCR3 1C7 LY117\\
			\small P05362  & \small  ICAM1    & \small  P02730  & \small  SLC4A1 AE1 DI EPB3\\
			\small P11277  & \small  SPTB SPTB1  & \small  Q08495  & \small  DMTN DMT EPB49\\
			\small P11413  & \small  G6PD  & \small  Q16570  & \small  ACKR1 DARC FY GPD\\
			\small P16157  & \small  ANK1 ANK  & \small  Q8TCT6  & \small  SPPL3 IMP2 PSL4\\
			\small P16284  & \small  PECAM1  & \small  Q8TCT7  & \small  SPPL2B IMP4 KIAA1532 PSL1\\
			\small Q99836  & \small  MYD88  & \small  P01375  & \small  TNF TNFA TNFSF2 \\
\hline

		\end{tabular}
	\end{minipage}
\end{table}

\subsection{Similarity Analysis}\label{subsec5}
The GO term ``semantic similarity analysis" is used to group related genes. A few well-known similarity approaches were applied to the dataset in order to determine the optimum similarity approach to be used on the gene ontology data. We are following standard procedures for all similarity tests:
\begin{enumerate}
	\item  The same data resource was used for all similarity tests.
	\item  If the similarity value exceeds the criterion of $0.60$ (or $60$\%), then similarity is $1$, else it is $0$.
	\item  Two-dimensional matrices with similarity values of $1$ and $0$ between the genes were created.
\end{enumerate}
\subsubsection{Cosine similarity}\label{subsec6}
The similarity between two numerical sequences can be calculated by using the cosine metric. Cosine similarity (${similarity}_c$), or the cosine of the angle between the vectors, is calculated by dividing the dot product of the vectors by their product. In a space called the inner product, the sequences are regarded as vectors. As a result, the cosine similarity only takes into account the angle of the vectors and not their magnitudes. The cosine similarity resides in the $[-1, 1]$ range. Cosine similarity has the benefit of being simple, especially for sparse vectors where just the non-zero coordinates need to be taken into account. Using CountVectorizer or TfidfVectorizer (which also provides frequency counts for each gene) supplied by SciKit Learn, we first determine the word count in the sentence in Python before using the cosine similarity algorithm. A Pandas dataframe or sparse matrix can be used as inputs for this method. The output is a matrix of similarity values as a result. A vector with two non-zero values can be obtained by using the Euclidean dot product formula: $C(A \cdot B) = ||X||\hspace{1mm} ||Y||\cdot \cos \theta $. ${similarity}_c$ between two n-dimensional attribute vectors $X$ and $Y$ is represented using a dot product and magnitude as: 
\begin{equation}
	C_S(X \cdot Y) = cos\theta = \frac{X \cdot Y}{||X|| \hspace{1mm} ||Y||} = \frac{\sum_{i = 1}^{n} X_i Y_i}{  \sqrt{\sum_{i=1}^{n}}X_{i}^2 \sqrt{\sum_{i=1}^{n}}Y_{i}^2    } , \label{cs}
\end{equation}
where $X_i$ and $Y_i$ are components of vector $X$ and $Y$, respectively. The resulting similarity spans from $0$ indicating orthogonality or decorrelation to $1$ suggesting precisely the same, with in-between values denoting intermediate similarity or dissimilarity. The resultant similarity can be expressed as a ratio between $-1$ and $1$, where $1$ means exactly the same.
\begin{figure}[htbp]
	\begin{minipage}[b]{1\textwidth}
		\centering
		\includegraphics[width=12cm, height=5cm]{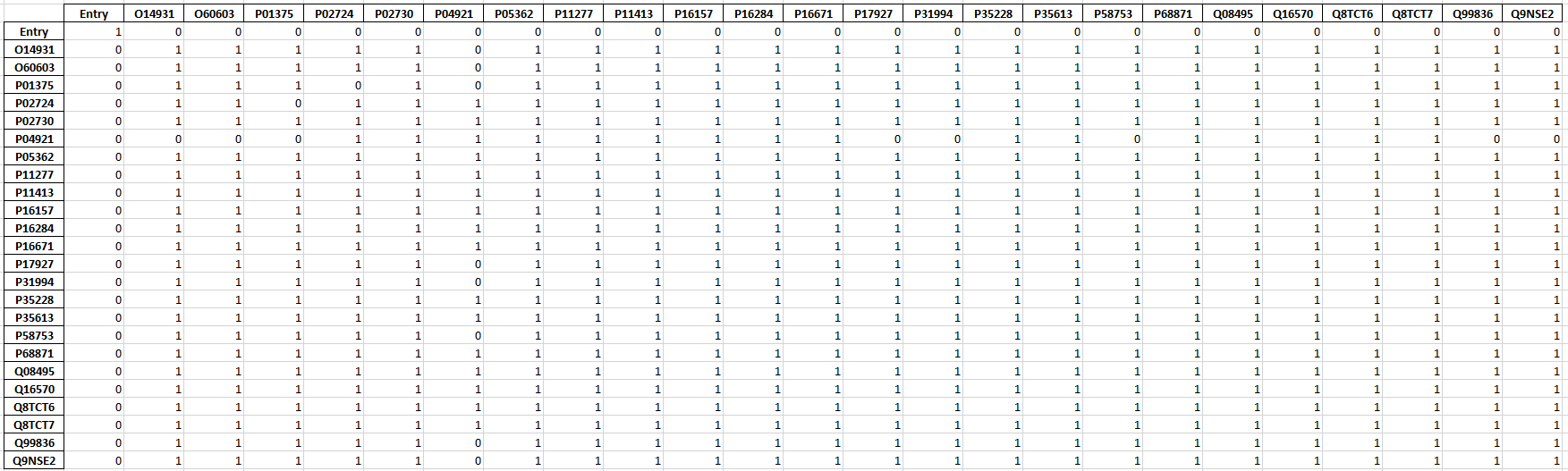}
		\caption{The Cosine similarity score}\label{hub_cs}
	\end{minipage}
\end{figure}
\subsubsection{Jaccard similarity coefficient}\label{subsec7}
A statistic for evaluating the diversity and similarity of sample sets is the Jaccard index \cite{jaccard1912distribution}, commonly referred to as the Jaccard similarity coefficient \cite{murphy1996finley}. The size of the intersection divided by the size of the union of the sample sets defines the Jaccard coefficient, which assesses similarity between finite sample sets:
\begin{equation}
	J_S(J_X, J_Y) = \frac{|J_X \cap J_Y|}{|J_X \cup J_Y|} = \frac{|J_X \cap J_Y|}{ |J_X| + |J_Y| - |J_X \cap J_Y| },  \label{js}
\end{equation}
Be aware that $0\le J_S(J_X, J_Y) \le 1$ exists by purpose. $J_S(J_X, J_Y) = 0$ if $J_X$ intersection $J_Y$ is empty. In fields like computer science, ecology, genetics, and other studies that work with binary or binarized data, the Jaccard coefficient is frequently employed. This allows us to compare two gene annotations for similarities. ``0" and ``1" are two values that are used to express degrees of similarity. Values of ``0" and ``1" indicate differences between the two sets of gene data. Each gene's corpus Jaccard similarity only contains one specific group of genes. Lemmatization is first used to condense gene data into a single root word, which is then used to calculate Jaccard similarity, which measures similarity.
\begin{figure}[htbp]
	\begin{minipage}[b]{1\textwidth}
		\centering
		\includegraphics[width=12cm, height=5cm]{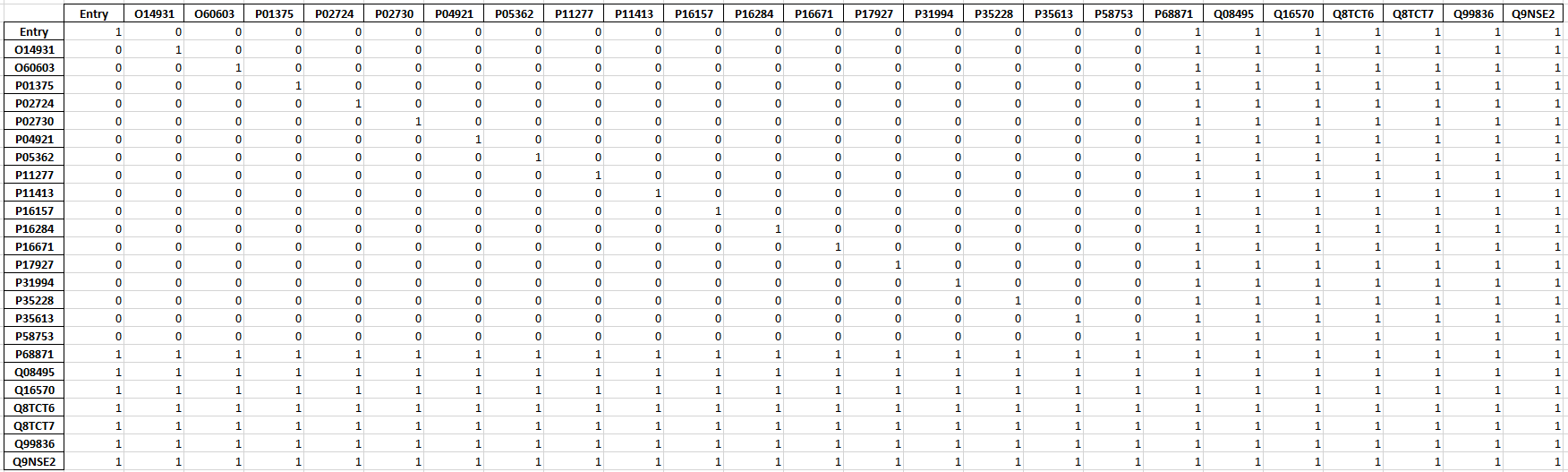}
		\caption{The Jaccard similarity score}\label{hub_js}
	\end{minipage}
\end{figure}

\subsubsection{Levenshtein ratio and distance measure}\label{subsec8}
An information theory and computer science metric called Levenshtein distance ($L_d$) is used to compare two sequences. Here, one word is transformed into another by adding, deleting, or substituting single characters \cite{levenshtein1966binary}. Strings are used to demonstrate the$L_d$ with an unequal length. 
$L_d$ between two strings can range from ``0" to ``1". The Levenshtein distance employs dynamic programming techniques like spell-checking and string matching. Let $Lev(L_X, L_Y)$ be the Levenshtein distance between two gene-annotated data of lengths $|L_X|$ and $|L_Y|$, respectively. Then conditionally we may represent as:
\begin{table}[H]%
	
	\begin{minipage}{\columnwidth}
		\caption{Levenshtein ratio and distance measure}\label{lev3}%
		\begin{tabular}{@{}c@{\hskip 2.6cm}c@{}}
			\hline
			Condition (if)  & Value \\
   \hline
			$L_X$ = 0 & $|L_Y|$ \\
			$L_Y$ = 0 & $|L_X|$ \\
			$L_X[0] = L_Y[0]$ & $Lev(tail(L_X), tail(L_Y))$\\
			otherwise & 1 + min \{$l_1$,$l_2$, $l_3$ \}  \\
			\hline
		\end{tabular}
	\end{minipage}
		
\end{table}

where $s[n]$ is the string's $n^{th}$ character (counting from $0$), and $s[n]$ is the tail of some string $s$, where the tail of some string is a string made up of all characters except for the first.
\begin{figure}[htbp]
	\begin{minipage}[b]{1\textwidth}
		\centering
		\includegraphics[width=12cm, height=5cm]{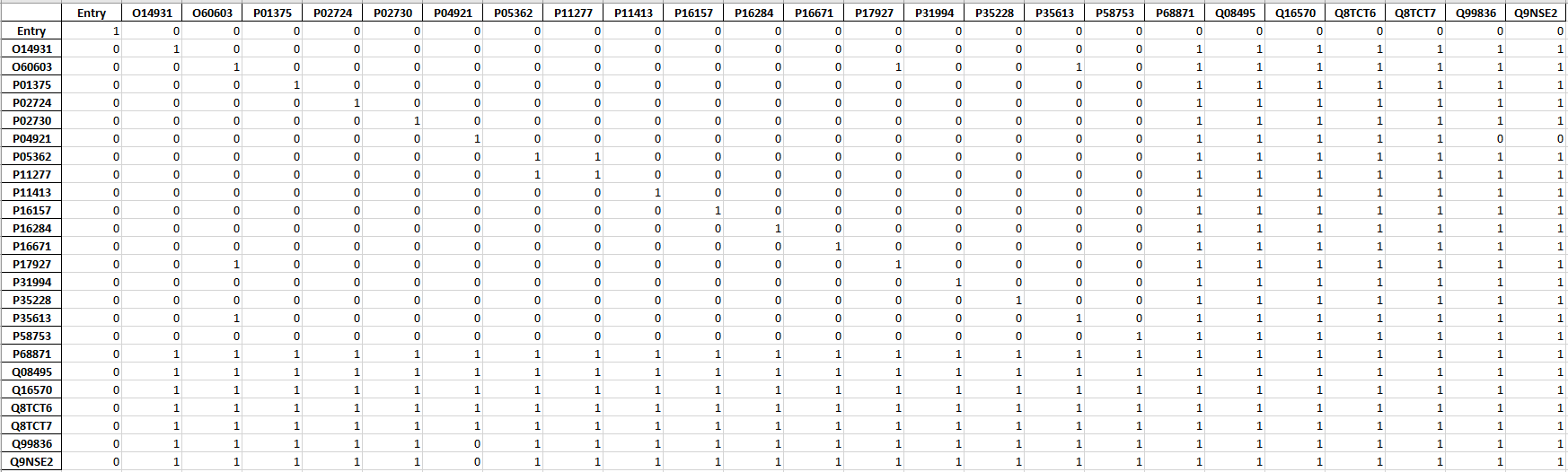}
		\caption{The Levenshtein ratio and distance measure similarity score}\label{hub_l}
	\end{minipage}
\end{figure}

\subsubsection{Pairwise document similarity}\label{subsec9}
A textual document similarity method based on the weights of terms in each document and the common terms (information) shared by two documents is known as the pairwise document similarity method (PDSM) \cite{oghbaie2018pairwise}. A weighting method determines a term's weight, which represents its importance in the document. The TF-IDF (Term Frequency-Inverse Document Frequency) form was utilised to calculate pairwise similarity. We have used the Scikit Learn TfidfVectorizer module to implement PDSM. Pairwise document similarity is defined by: 
\begin{equation}
	PDSM(X, Y) = \left(\frac{X \cap Y}{X \cup Y} \right) \times \frac{PF(X, Y) + 1}{M - AF(X, Y) + 1}, \label{pdsm}
\end{equation} 
The following formula is used to determine the intersection $({X \cap Y} = \sum_{1}^{M} Min(w_{xi}, w_{yi}) \label{inter})$ and union $({X \cup Y} = \sum_{1}^{M} Max(w_{xi}, w_{yi}) \label{union})$ of two documents (where $w_{ji} > 0$ denotes the $i^{th}$ term weight in document $j$). The number of phrases that are present and those that are absent are denoted by $PF(d1, d2)$ and $AF(d1, d2)$, respectively. $1$ is added to the numerator and denominator in order to prevent a Divide-by-Zero error.
\begin{figure}[H]
	\begin{minipage}[b]{1\textwidth}
		\centering
		\includegraphics[width=12cm, height=5cm]{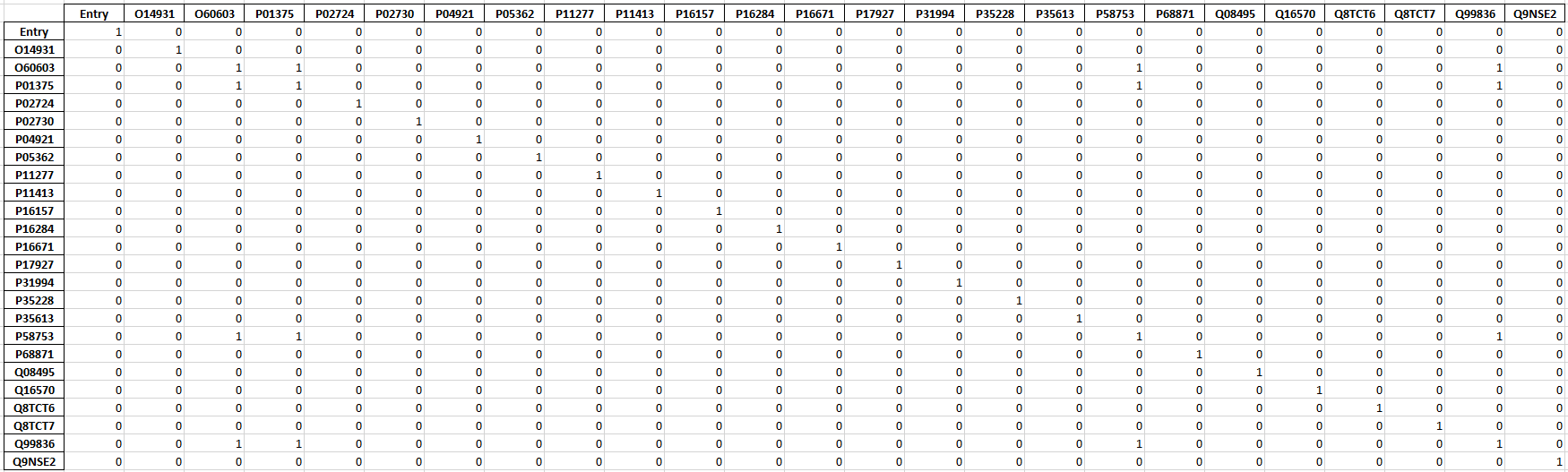}
		\caption{The Pairwise document similarity score}\label{hub_ps}
	\end{minipage}
\end{figure}

\paragraph{Conclusion:}
The methodology with the highest number of $1's$ was determined to be the best method for similarity analysis based on the output of similarity values ($1's$ and $0's$), as more $1's$ would indicate more comparable genes. In our work, the Cosine Similarity metric yielded the most similar values in relation to the data resource employed. As shown in Fig. \ref{hub_js} to \ref{hub_l} which are the results obtained, the similar genes have a value of $1$, while the dissimilar genes have a value of $0$.
\subsection{Protein Protein Interaction Network}\label{subsec2}
PPIs play an essential role in almost every cell process, so understanding their function in normal and disease states is crucial \cite{das2023analyzing}. PPI is important in predicting target protein protein function and molecule drug ability. As a set of interactions, the majority of genes and proteins realise phenotype functions. A PPIN is a mathematical representation of a protein's physical interaction with its environment \cite{das2022analytical}. GO data of Malaria has been used to test our criteria for defining protein interaction hubs.

\subsection{Centrality Analysis}\label{subsec12}
This section examines network centrality metrics, which we employ to pinpoint nodes (proteins) of structural significance. The term ``centrality" initially referred to a node's position within a network's hierarchy. Its topological roots have been abstracted as a phrase, and it now very broadly refers to how significant nodes are to a network. Although topological centrality has multiple operationalizations, it has a precise definition. On the other hand, there are numerous operationalizations and meanings of ``importance" in the network. Here, we'll look at various operationalizations and interpretations of centrality with page rank. DC, BC, CC, and EC are the four widely used centrality measurements; each has advantages and disadvantages. Apart from these four centralities, we have also mentioned load centrality and page rank for better results.
The term ``centrality" describes how crucial a node or edge is to the network's connection or movement within a graph. We used centrality measurements to determine the hub or powerful node in the PPI network that was built. It is crucial to identify hub nodes since they will be connected to the majority of other nodes in the network and exert a significant influence over all the others. It is possible to think of the hub node's functionality as the network's overall functionality. Let assume that $G(V, E)$ is a graph with $|V|$ vertices and $|E|$ edges. Suppose $A = (a_{v,u})$ be the adjacency matrix. If vertex $v$ is linked to the vertax $u$ then $a_{v,u} = 1$ otherwise $a_{v,u} = 0$. 
A partial output of each centrality measure based on the data resource is given in each centrality section.

\subsubsection{Degree Centrality}\label{subsubsec1}
Degree centrality is one of the easiest centrality measurements to calculate. The number of edges incident to a vertex in a graph, counted twice using loops, is known as the vertex's degree. With regard to the data resources needed, this is a portion of the near centrality measure's output.
\begin{equation}
	C_d\mathit(v) = \frac{deg\mathit(v)}{\mbox{max deg}_{u \in \mathit{v}} \mathit(u)}  .\label{hub_dc}
\end{equation}

Degree centrality ranges from $0$ to $1$, and a value near 1 indicates that the node is likely to have a maximum degree. Nodes in a network can be ranked according to their degree centrality in order to determine those that are the most prominent or influential.


\subsubsection{Closeness Centrality}\label{subsubsec2}
Closeness centrality (CC) determined as the total length of the shortest paths connecting it to every other node in the graph, is a measure of a node's centrality in a network. How near a node is to every other node in the network is indicated by its centrality. This is a portion of the closeness centrality measure's result in relation to the data resource that was used.    
\begin{equation}
	C_c\mathit(v) = \frac{\lvert V \rvert  - 1}{\sum_{u\in V - {\mathit{\{ v\}}}}^{}d\mathit (u, v)}  
\end{equation}

A node's number is represented by $|V|$ and its distance from another node is indicated by $d(u, v)$ ($u$ and $v$ are two different node). A node with a high CC value is considered to be of higher quality. Epidemic modeling uses the measure to examine or restrict disease spread.


\subsubsection{Betweenness Centrality}\label{subsubsec3}
A node's importance is determined by its betweenness centrality (BC). The number of edges the path passes through, the total of the weighted edges, or every pair of vertices with at least one shortest path between them.

\begin{equation}
	C_b\mathit(v) =\sum_{xy \in V - \mathit{\{v\}}}^{} \frac{\sigma_{xy} (\mathit{v})}{\sigma_{xy}} \label{hub_bc}
\end{equation}

where the frequency of shortest paths in the network between nodes $x$ and
$y$ is indicated by $\sigma_{xy}$ and $\sigma_{v}$ denotes the same passing through $v$. If $x = 1$, then $\sigma_{xy} = 1$. An epidemiological analysis of disease spreading can benefit from the BC by identifying super spreaders.


\subsubsection{Eigenvector centrality}\label{subsubsec4}
Eigenvector centrality (ER), often known as eigen centrality, is a metric for a node's power within a network. A node will have a high eigenvector centrality if it is directed by a large number of other nodes.
The Eigenvector centrality of vertex $v$ can be defined as: 

\begin{equation}
	x_(v) = \frac{1}{\lambda} \sum_{u\in M(v)}^{} x_u = \frac{1}{\lambda} \sum_{u\in V}^{}a_{v, u}x_{u}     \label{hub_ec} 
\end{equation}

Where $M_(v)$ is the set of neighbors of $v$ and $\lambda$ is a constant. With a small rearrangement, this can be rewritten in vector notation as the eigenvector equation: $ Ax = \lambda x$. In general, a non-zero eigenvector solution will exist for a wide range of various eigenvalues lambda. However, the Perron-Frobenius theorem \cite{pillai2005perron} indicates that only the largest eigenvalue produces the desired centrality measure due to the extra requirement that all items in the eigenvector be non-negative.

\begin{equation}
	C_l = \sum_{x,y \in V}^{} \sigma_{x,y}(v)
\end{equation}

Where typically, it is assumed that $\sigma_{x,y} = 1$ and that $x \notin y$, $x \notin v$, $y \notin v$.

\subsubsection{Page Rank}\label{subsubsec6}
The Page Rank (PR)\cite{ivan2011web}  algorithm ranks web content by looking at how links between sites link to each other. Protein interaction networks, as well as any other type of network, can be use it. It uses random walks to identify individuals who are commonly encountered along such walks. Those individuals are viewed as central. Mathematically, it can be defined as:
\begin{equation}
	C_{PR}\mathit(v_i) = \frac{1-d}{\vert V \vert} + d \sum_{\mathit(v_t)\in Inneighbor\mathit(v_i)}^{} \frac{C_{PR}(\mathit{v_t})}{outdeg(\mathit{v_t})}
\end{equation} 

A damping factor called $d$ is considered a constant value, and is usually defined as $0.85$.

\subsection{Clustering}
The degree to which nodes in a graph tend to cluster together is quantified by the clustering coefficient (CCo). The transitivity of a graph has a direct impact on its clustering coefficient. Let's we are computing the Clustering Coefficients (CCo) for our  unweighted graphs $G$, the clustering of a node $x$ is the fraction of possible triangles through that node that exist:
\begin{equation}
	{CCo}_x = \frac{2T(x)}{deg(x)(deg(x-1))}     \label{cco}
\end{equation}
where $T(x)$ is the number of triangles through node $x$ and $deg(x)$ is the degree of $x$.
\section{Results}\label{sec5}
\begin{figure*}
	
	\begin{minipage}[b]{.5\linewidth}
		\centering
		\includegraphics[width=5cm, height=3cm]{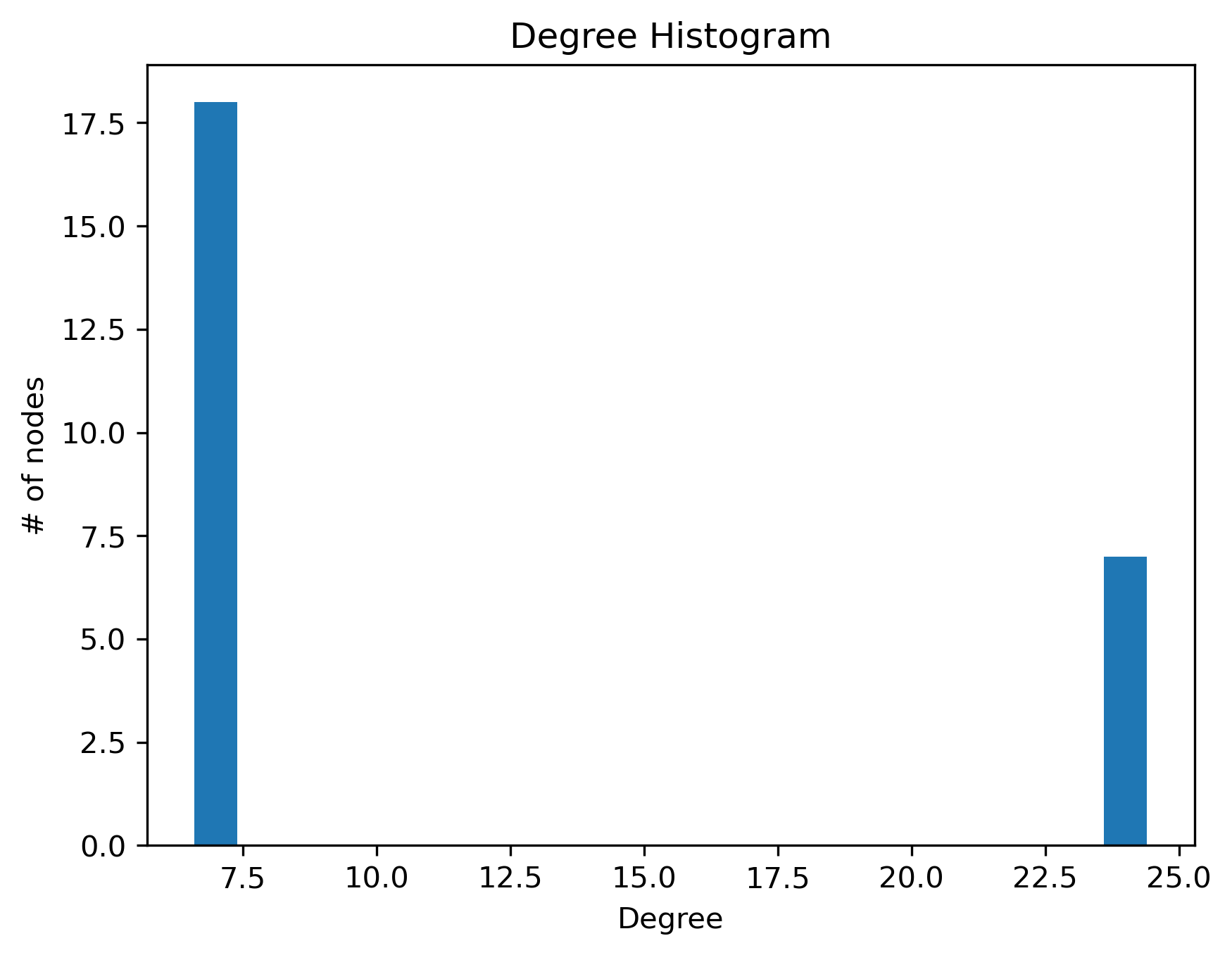}
		\caption{DD of network \ref{random_jaccard_ppi} }\label{node_histo_js}
	\end{minipage}
	\begin{minipage}[b]{.5\linewidth}
		\centering
		\includegraphics[width=5cm, height=3cm]{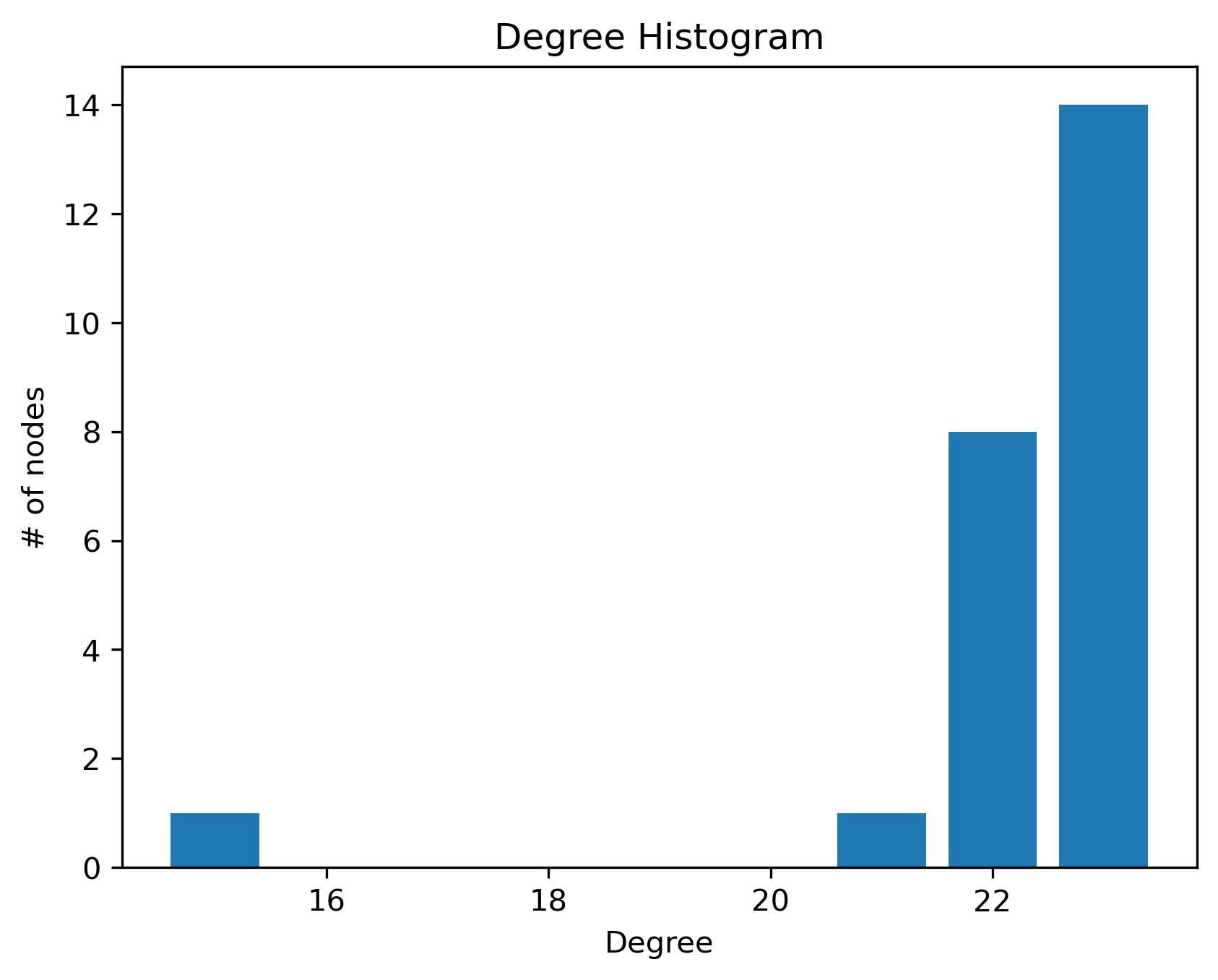}
		\caption{DD of network \ref{random_cs_ppi}}\label{node_histo_cosine}
	\end{minipage}
	\begin{minipage}[b]{.5\linewidth}
		\centering
		\includegraphics[width=5cm, height=3cm]{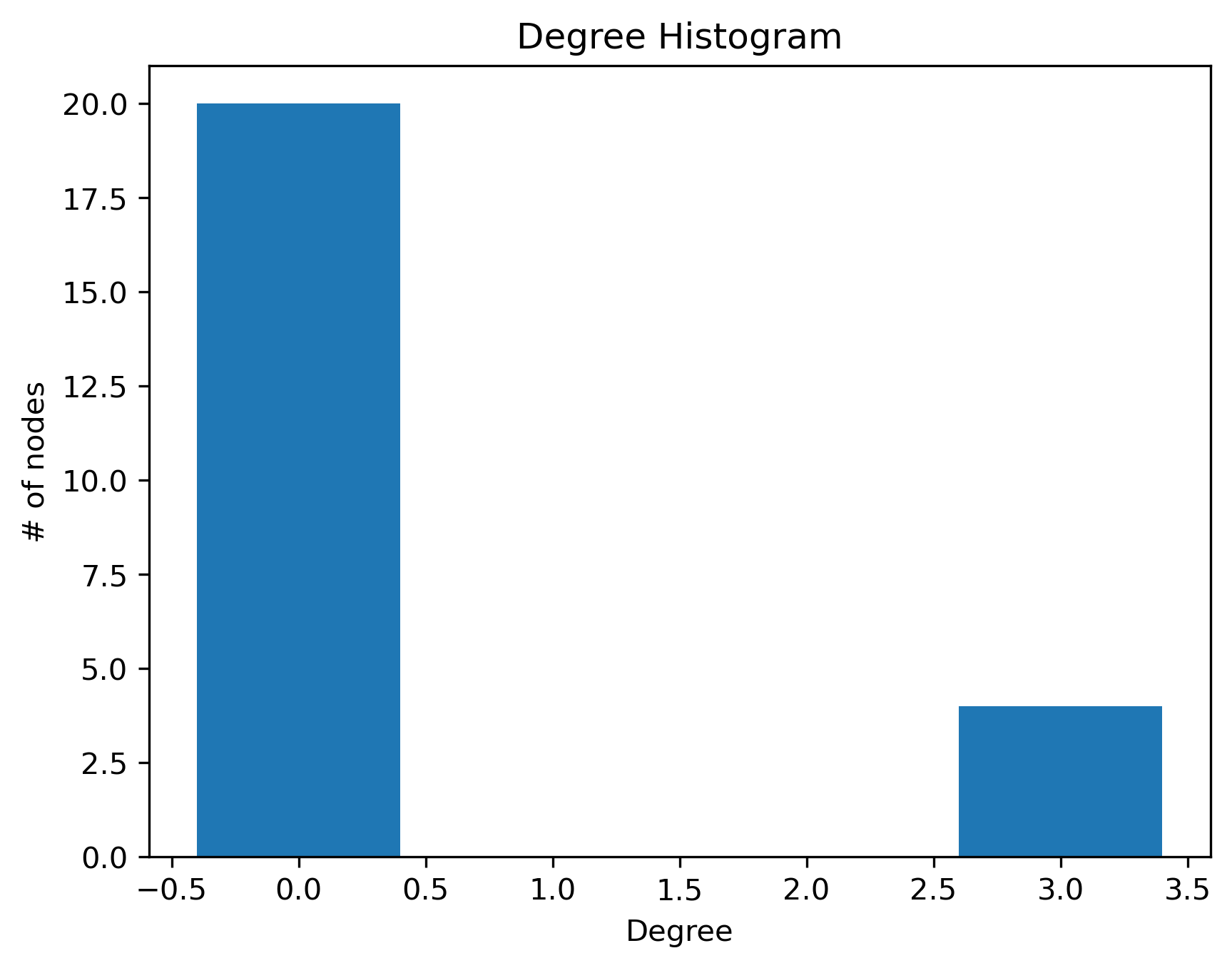}
		\caption{DD of network \ref{random_ps_ppi}}\label{node_histo_ps}
	\end{minipage}
	\begin{minipage}[b]{.5\linewidth}
		\centering
		\includegraphics[width=5cm, height=3cm]{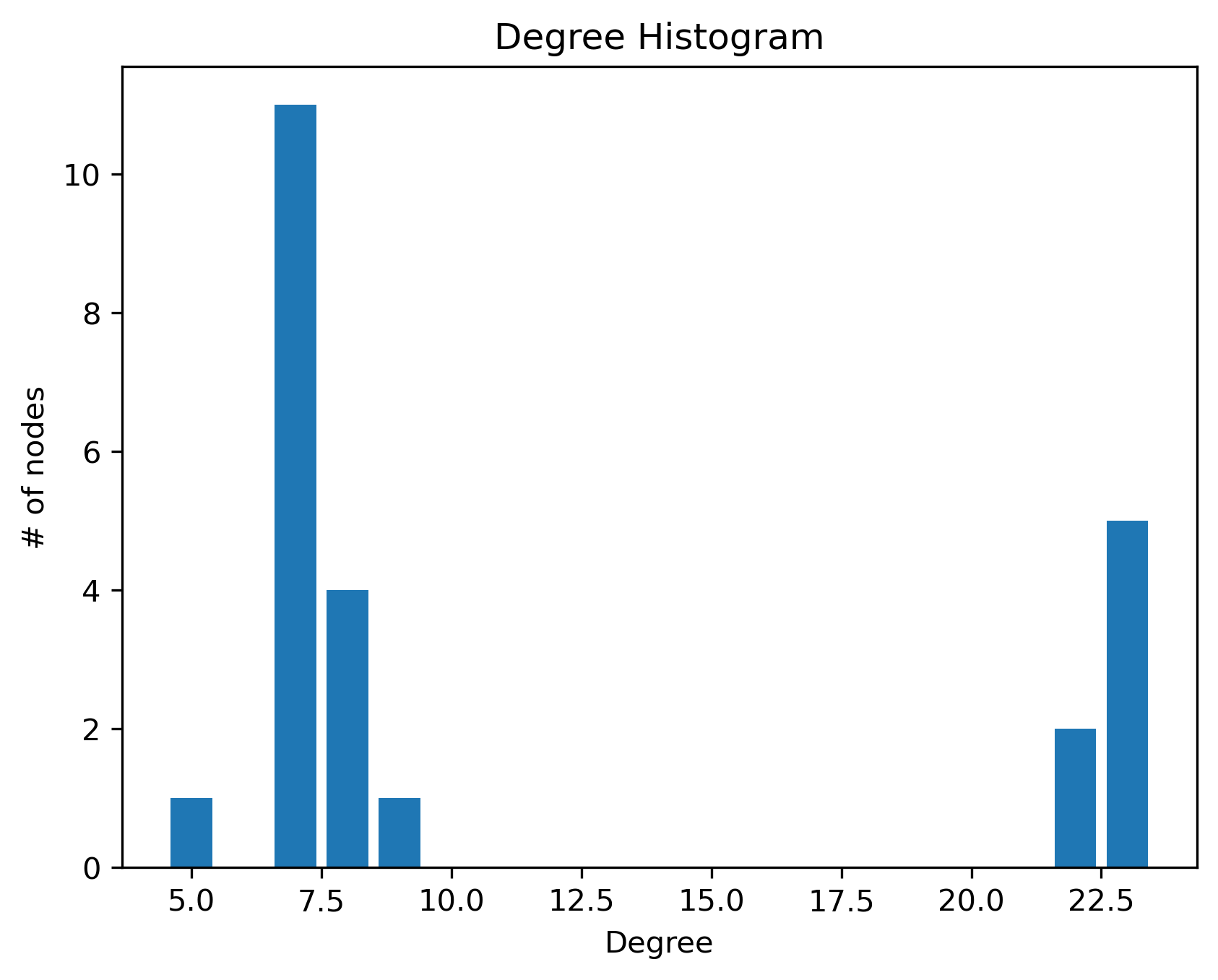}
		\caption{DD of network \ref{random_l_ppi} }\label{node_histo_ls}
	\end{minipage}
\end{figure*}

A network of nodes made up of comparable genes can be created, with the nodes representing the genes and the edges between the nodes only being constructed if the similarity weight is $1$. Similar genes were gathered into one network based on the findings of the different similarity measures. In order to gain a better understanding of a network, a variety of network analyses can be conducted. There are several topological features that can be found in network topology, such as degree distribution (Fig. \ref{node_histo_cosine} to \ref{node_histo_ls}), diameter ($N_1, N_2, N_3: 2$), and the clustering coefficient (Table \ref{tab3}) of interaction networks. An indicator of the relationship between a node's neighbours is its clustering coefficient, which ranges from $1$ to $0$. The global properties of the malaria GO network are shown in table \ref{tab3}. All four network except $N_4$ has an average node degree greater than $11$. The $3$ network ($N_1, N_2, N_3$) has density $\ge 0.49$. The highest density is $0.967$ ($N_1$) and the lowest density is $N_4$ ($0.022$). The average LCC is pretty good (highest $0.977$). 

The centrality scores of $N_1$ are provided in Table \ref{tab4}, which enables us to determine the significance of the protein. The network's average Local Clustering Coefficient (LCC) is $0.977$ and its maximum degree is $23$. Fig. \ref{node_histo_cosine} displays the degree distribution. The LCC represent the density of connections among neighbours and vary from $0$ to $1$. Nodes with higher values are part of clusters that are closely related. If a node has a value of $1$, it is regarded as a member of the clique. Because they are a part of the clique, the proteins $P01375$ and $P04921$ in Table reftab4 possess CCo value as $1$. There were $9$ hub proteins found among $24$ hub proteins namely `P11413', `P16284', `P16671', `P35228', `P68871', `Q08495', `Q16570', `Q8TCT6', `Q8TCT7' and these are visualized in Fig. \ref{hub_node} with red colour. We have compared the hub node getting from the proposed approach with the vote rank algorithm in Table \ref{tab5}. We can see the annotation cluster in Table \ref{tab6} which contains 4 clusters.
\subsubsection{VoteRank Algorithm}\label{subsec13}
\begin{figure*}
	\begin{minipage}[b]{1\linewidth}
		\centering
		\includegraphics[width=8cm, height=6cm]{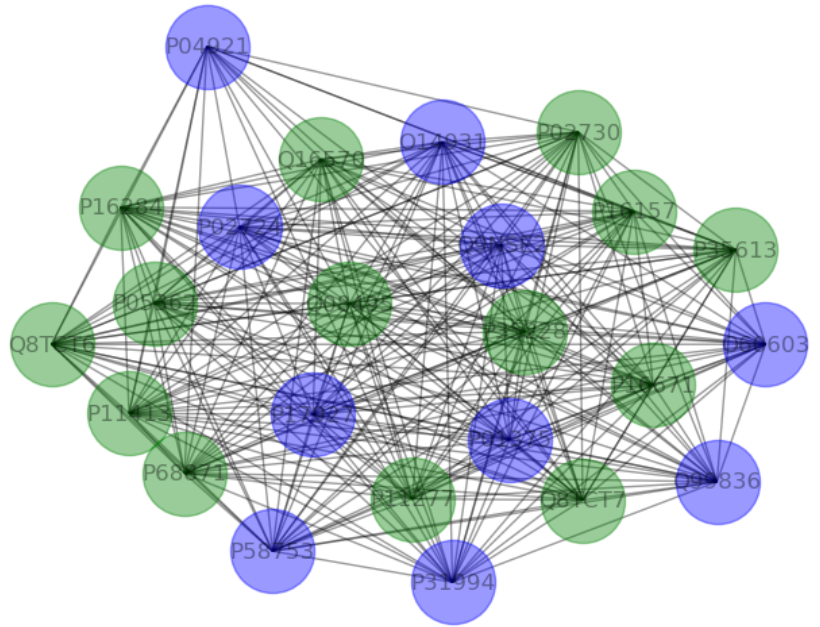}
		\caption{The hub node (green node) from voterank}
		\label{voterank}
	\end{minipage}
	\hspace{1cm}
	\begin{minipage}[b]{1\linewidth}
		\centering
		\includegraphics[width=8cm, height=6cm]{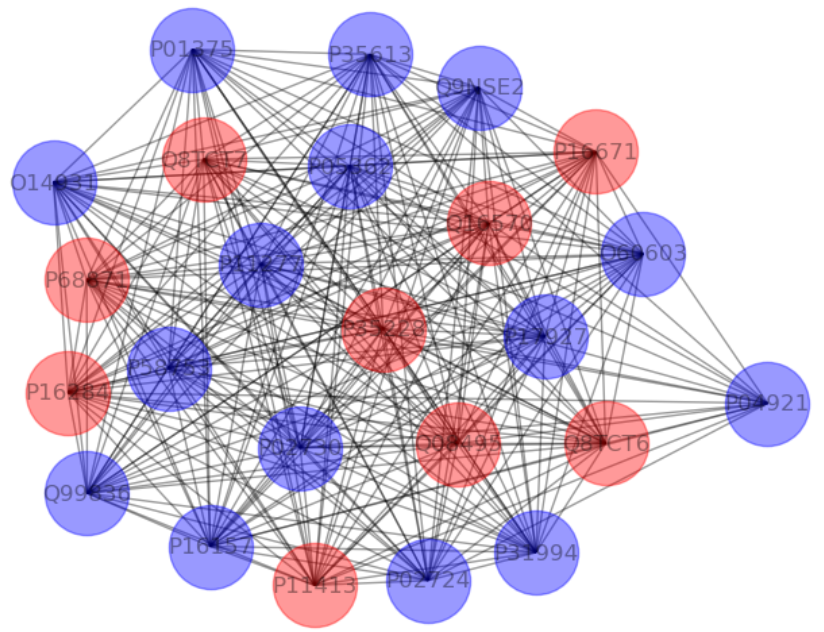}
		\caption{The hub node (red node) from proposed approach}
		\label{hub_node}
	\end{minipage}
	\hspace{1cm}
	\begin{minipage}[b]{1\linewidth}
		\centering
		\includegraphics[width=8cm, height=6cm]{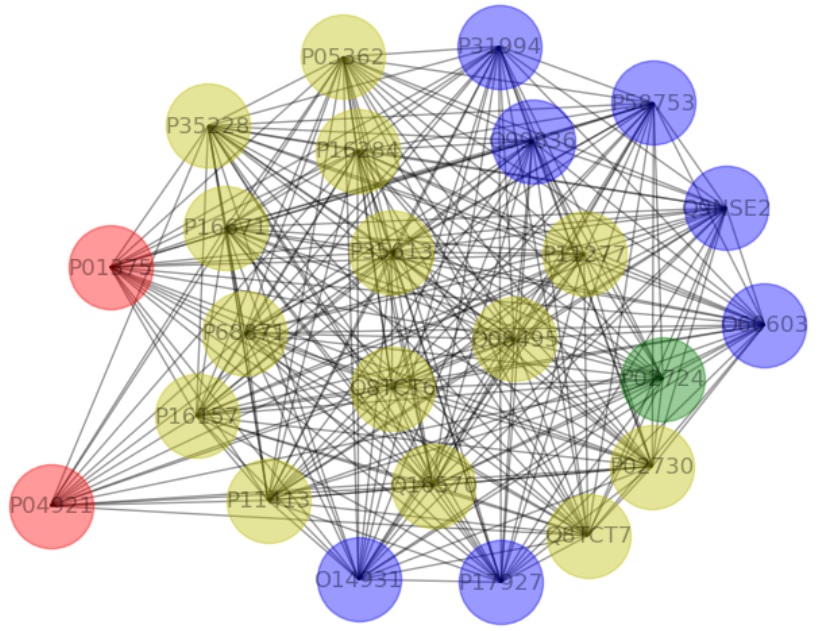}
		\caption{4 clusters (denoted by red, green, blue, and yellow)}
		\label{cluster}
	\end{minipage}
\end{figure*}

VoteRank \cite{zhang2016identifying} uses a voting system to determine the order of the nodes in a graph $G$. In the actual world, if person M has assisted person N, M's ability to support others would typically wane. This article presents VoteRank, a vote-based method for identifying influential spreaders, from this point of view. The basic goal of VoteRank is to select a group of spreaders one at a time in accordance with the voting scores that nodes receive from their neighbours. Here, we have got $14$ influential spreaders according to the vote ranking algorithm namely `P35228', `Q8TCT6', `P35613', `P68871', `P16157', `P16671', `P11277', `Q16570', `Q8TCT7', `P16284', `P02730', `Q08495', `P05362', and `P11413'. The protein is visualized in green colour in Fig. \ref{voterank}. To ensure that the hub node that had been constructed was the same across all four centrality measurements, we used four distinct metrics.

\begin{table}[H]%
	\begin{minipage}{\textwidth}
		\caption{Global properties of five network}\label{tab3}%
		\begin{tabular}{@{}cccccc@{}}
			\hline
			S.A Type  & \#Edge & Max. Degree & Avg. Node Degree & Density & Avg. LCC \\
			\hline 
			\ref{subsec6} & 267 & 23 & 22.25 & 0.967 & 0.977 \\ 
			\ref{subsec7} & 147 & 24 & 12.25 & 0.49  & 0.845 \\
			\ref{subsec8} & 141 & 23 & 11.75 & 0.511 & 0.845\\
			\ref{subsec9} & 6   & 3  & 0.5   & 0.022 & 0.167\\
			\hline
		\end{tabular}
	\end{minipage}
\end{table}

\begin{table}[H]%
	
	\begin{minipage}{\columnwidth}
		\caption{Centrality measure and some important score}\label{tab4}%
		\begin{tabular}{c@{\hskip 7mm}c@{\hskip 7mm}c@{\hskip 7mm}c@{\hskip 7mm}c@{\hskip 7mm}c@{\hskip 7mm}c@{\hskip 7mm}}
			\hline
			\textbf{Entry}	&	\textbf{DC} & \textbf{CC}	&	\textbf{BC}	&	\textbf{EC}	&	\textbf{PR} & \textbf{CCo}\\ 
			\hline
			
			O14931 	&	0.957	&	0.958	&	0.000	&	0.203	&	0.041	&	0.996		\\
			O60603 	&	0.957	&	0.958	&	0.000	&	0.203	&	0.041	&	0.996		\\
			P01375 	&	0.913	&	0.920	&	0.000	&	0.195	&	0.040	&	1.000		\\
			P02724 	&	0.957	&	0.958	&	0.002	&	0.201	&	0.041	&	0.970		\\
			P02730 	&	1.000	&	1.000	&	0.002	&	0.209	&	0.043	&	0.964		\\
			P04921 	&	0.652	&	0.742	&	0.000	&	0.140	&	0.030	&	1.000		\\
			P05362 	&	1.000	&	1.000	&	0.002	&	0.209	&	0.043	&	0.964		\\
			P11277 	&	1.000	&	1.000	&	0.002	&	0.209	&	0.043	&	0.964		\\
			P11413 	&	1.000	&	1.000	&	0.002	&	0.209	&	0.043	&	0.964		\\
			P16157 	&	1.000	&	1.000	&	0.002	&	0.209	&	0.043	&	0.964		\\
			P16284 	&	1.000	&	1.000	&	0.002	&	0.209	&	0.043	&	0.964		\\
			P16671 	&	1.000	&	1.000	&	0.002	&	0.209	&	0.043	&	0.964		\\
			P17927 	&	0.957	&	0.958	&	0.000	&	0.203	&	0.041	&	0.996		\\
			P31994 	&	0.957	&	0.958	&	0.000	&	0.203	&	0.041	&	0.996		\\
			P35228 	&	1.000	&	1.000	&	0.002	&	0.209	&	0.043	&	0.964		\\
			P35613 	&	1.000	&	1.000	&	0.002	&	0.209	&	0.043	&	0.964		\\
			P58753 	&	0.957	&	0.958	&	0.000	&	0.203	&	0.041	&	0.996		\\
			P68871 	&	1.000	&	1.000	&	0.002	&	0.209	&	0.043	&	0.964		\\
			Q08495 	&	1.000	&	1.000	&	0.002	&	0.209	&	0.043	&	0.964		\\
			Q16570 	&	1.000	&	1.000	&	0.002	&	0.209	&	0.043	&	0.964		\\
			Q8TCT6 	&	1.000	&	1.000	&	0.002	&	0.209	&	0.043	&	0.964		\\
			Q8TCT7 	&	1.000	&	1.000	&	0.002	&	0.209	&	0.043	&	0.964		\\
			Q99836 	&	0.957	&	0.958	&	0.000	&	0.203	&	0.041	&	0.996		\\
			Q9NSE2 	&	0.957	&	0.958	&	0.000	&	0.203	&	0.041	&	0.996		\\
			
			\hline
		\end{tabular}
	\end{minipage}
	
\end{table}


\begin{table}[H]
	
	\begin{minipage}{\textwidth}
		\caption{Comparing hub node of Voterank and proposed approach}\label{tab5}%
		\begin{tabular}{@{}ll@{}}
			\hline
			\tiny VoteRank  & \tiny `P11413', `P16284', `P16671', `P68871', `Q08495', `P35228', `Q16570', `Q8TCT6', `Q8TCT7',\\ & \tiny  `P35613', `P16157', `P11277', `P02730', `Q08495', `P05362'\\
			\tiny Proposed approach & \tiny `P11413', `P16284', `P16671', `P35228'	`P68871', `Q08495'	`Q16570', `Q8TCT6'	`Q8TCT7'  \\ 
			
			\hline
		\end{tabular}
	\end{minipage}
\end{table}

\begin{table}[H]%
	
	\begin{minipage}{\textwidth}
		\caption{Annotation cluster}\label{tab6}%
		\begin{tabular}{l@{\hskip 1mm}l@{\hskip 1mm}l@{\hskip 1mm}}
			\hline
			\tiny  & \tiny count &\tiny	cluster (Bold entry denotes the hub node)\\
			\hline
			\tiny \textcolor{green}{$C_1$} & \tiny 1 & \tiny `P02724' \\
			
			\tiny \textcolor{red}{$C_2$} & \tiny 2 &\tiny `P01375',  `P04921' \\
			
			\tiny \textcolor{blue}{$C_3$} &\tiny 7 & \tiny `O14931',  `O60603', `P17927',  `P31994', `P58753',  `Q99836',  `Q9NSE2' 
			\\

			\textcolor{yellow}{$C_4$} &\tiny 14 & \tiny `P02730',  `P05362',  `P11277',  \textbf{`P11413'},  `P16157', \textbf{`P16284'}, \textbf{`P16671'},  \textbf{P35228}, P35613,  \textbf{P68871},  \textbf{Q08495}, \\& &\tiny \textbf{Q16570},  \textbf{Q8TCT6},  \textbf{Q8TCT7} 
			\\
			
			\hline
		\end{tabular}
	\end{minipage}
\end{table}
\begin{table}[H]
	\begin{minipage}{\textwidth}
		\caption{Gene function of Portion of data.\label{tab2}}
		\tabcolsep=0pt
		\begin{tabular}{@{\extracolsep{\fill}}ll@{\extracolsep{\fill}}}
			\hline
			
			\tiny Gene & \tiny Functions \\
			\hline
			\tiny NOS2 &  \tiny This molecule serves as a messenger throughout the body by producing nitric oxide.\\ 
			
			\tiny SPPL3 & \tiny LT domain borders of type II membrane protein substrates are where I-CLiP cleaves those proteins. \\ 
			
			\tiny NCR3 & \tiny Stimulates NK cell cytotoxicity against nearby cells and by controlling NK, for instance, tumor cells are destroyed. \\
			
			\tiny MYD88 & \tiny Signaling pathway involving Toll-like receptors and IL-1 receptors in the innate immune system. \\  
			
			\tiny TNF &   \tiny Macrophages secrete it largely, and it is capable of causing tumor cell death\\
			
			\tiny FCGR2B  &  \tiny \hspace{2mm}In addition to phagocytosis, it modulates antibody  production by B cells.\\
			\hline
		\end{tabular}
	\end{minipage}
\end{table}
\subsection{Applications}\label{sec4}
Gene annotation is a representation of the gene's functional information. Sequence similarity and semantic similarity are correlated, which aids in predicting protein function. Genes with comparable expression patterns can be grouped together, which allows for further study. An important purpose of annotation is gene prediction, which aids in the investigation of a genome's protein binding sites. Any ontology approach's key drawback is the inability to employ partial GO annotation to cover any statistical data. A portion of database with functions are included in \ref{tab2}.

\section{Conclusion}\label{sec6}
In order to identify functionally related genes for the data resource employed, we applied and analysed semantic text similarity methods to obtain the best and most optimal similarity methods. Additionally, PPI networks of related genes were created, and various centrality measures were used to determine the hub nodes of the protein complexes. In order to classify and organise the genes and proteins into their appropriate groupings, this technique can be applied to a bioinformatics dataset. Further gene behaviour can be expected based on the traits of the discovered cluster group.

	\section{Compliance with Ethical Standards} 
	\begin{itemize}
		\item \textbf{Funding:} This is the work of the first author under her doctoral. This research received no external funding.
		\item \textbf{Disclosure of potential conflicts of interest:} On behalf of all authors, the corresponding author states that there is no conflict of interest.
		\item \textbf{Research involving human participants and/or animals:} This article does not contain any studies with human participants or animals performed by any of the authors.
	\end{itemize}
	


\section{Author Contributions}
Mamata Das designed and performed research; Mamata Das and K. Selvakumar analyzed data; Mamata Das wrote the paper; P.J.A. Alphonse read the paper.

\appendix


\bibliographystyle{elsarticle-num-names} 
\bibliography{160223_reference}





\end{document}